\documentclass[pra,showpacs,twocolumn]{revtex4}

\usepackage{amsmath,amssymb,amscd}
\usepackage{psfrag}

\usepackage{graphicx}

\usepackage[dvips]{epsfig}
\usepackage{epsfig}

\usepackage{bm}

\newcommand{\tabincell}[2]{\begin{tabular}{@{}#1@{}}#2\end{tabular}}

\usepackage{enumerate}
\usepackage{booktabs}
\usepackage{theorem}
\theorembodyfont{\rmfamily}
\newtheorem{theorem}{Theorem}

\newtheorem{corollary}{Corollary}

\theoremstyle{break}

\def\QED{~\rule[-1pt]{5pt}{5pt}\par\medskip}

\def\CalM{\mathcal{M}}

\def\Eq{Eq.~\eqref}
\def\Eqs{Eqs.~\eqref}

\DeclareMathOperator{\diag}{diag}

\DeclareMathOperator{\Var}{Var}

\DeclareMathOperator{\E}{\mathbb{E}}

\newcommand{\ma}[1]{\left[\begin{matrix} #1 \end{matrix}\right]}

\usepackage{color}

\begin{document}
\title{Quantum Secret Sharing with Continuous Variable Graph State}
\author{Yadong Wu$^{1}$, Runze Cai$^{1}$,
Guangqiang He$^{2,}$\footnote{Corresponding author. Email:
  gqhe@sjtu.edu.cn},
and Jun  Zhang$^{1,}$\footnote{Corresponding author. Email:
    zhangjun12@sjtu.edu.cn}}

\affiliation{$^1$Joint Institute of UM-SJTU, Shanghai Jiao Tong
  University, and Key Laboratory of System Control and Information
  Processing (Ministry of Education), Shanghai, 200240, China\\
  $^2$State Key Laboratory of Advanced Optical Communication Systems and
  Networks, and Department of Electronic Engineering, Shanghai Jiao
  Tong University, Shanghai, 200240, China}

\date{\today}

\begin{abstract}
  In this paper we study the protocol implementation and property
  analysis for several practical quantum secret sharing (QSS) schemes
  with continuous variable graph state (CVGS). For each QSS scheme, an
  implementation protocol is designed according to its secret and
  communication channel types. The estimation error is derived
  explicitly, which facilitates the unbiased estimation and error
  variance minimization. It turns out that only under infinite
  squeezing can the secret be perfectly reconstructed. Furthermore, we
  derive the condition for QSS threshold protocol on a weighted
  CVGS. Under certain conditions, the perfect reconstruction of the
  secret for two non-cooperative groups is exclusive, {\it i.e.} if
  one group gets the secret perfectly, the other group cannot get any
  information about the secret.
\end{abstract}


\maketitle

\section{introduction}
\label{sec:introduction}
Quantum cryptography provides a sophisticated approach to achieve the
communication security by taking advantage of quantum mechanics
principles~\cite{gisin2002quantum}. Among various schemes, quantum
secret sharing (QSS) is a general multi-partite information security
scheme that attracts extensive research
interests~\cite{PhysRevA.59.1829, PhysRevLett.83.648,
  PhysRevA.61.042311, PhysRevA.64.042311, PhysRevA.71.012328,
  PhysRevA.78.042309, PhysRevA.82.062315, PhysRevA.86.042303}.  It
allows one dealer to distribute a secret among a number of players in
such a way that a certain set of players can reconstruct the secret by
taking operations collaboratively and exchanging information.  In
contrast to quantum key distribution~\cite{bennett1984quantum} that
guarantees the secure communication between only two parties, QSS
enables multiple parties to communicate securely at the same time.

QSS has its origin in classical information theory. An early scheme
was given in~\cite{PhysRevA.59.1829} to share either classical or
quantum secret to three or four players by using the GHZ states.
Ref.~\cite{PhysRevLett.83.648} studied general threshold schemes to
share quantum secrets and showed that the quantum no-cloning theorem
is the only constraint on the existence of threshold schemes.
Ref.~\cite{PhysRevA.61.042311} further extended the results to general
access structures, including non-threshold schemes. These researches
have established theoretical foundations for many ensuing
investigations, {\it e.g.} hybrid schemes~\cite{PhysRevA.64.042311}
and twin-threshold schemes~\cite{PhysRevA.71.012328}.

On the other hand, graph state has been extensively studied in
applications such as quantum error
correction~\cite{PhysRevLett.101.090501, PhysRevA.78.012306,
  PhysRevA.78.042303, PhysRevA.79.042342}, entanglement
purification~\cite{PhysRevLett.91.107903, PhysRevA.71.012319,
  PhysRevA.74.052316}, entanglement
measurement~\cite{PhysRevA.69.062311, markham2007entanglement,
  Hajdusek2012entanglement}, and Bell
inequality~\cite{PhysRevLett.95.120405, PhysRevA.73.022303}.  In
recent years, the implementation of QSS with graph state was
introduced in~\cite{PhysRevA.78.042309, EPTCS9.10} to treat three
kinds of threshold QSS schemes in a unified graph state approach and
to propose embedded protocols in large graph states.
Ref.~\cite{PhysRevA.82.062315} generalized the results to prime
dimensions, and Ref.~\cite{PhysRevA.86.042303} investigated
non-threshold schemes. However, all these results are based on
discrete variable graph states.

Here we are interested in QSS with continuous variable graph state
(CVGS). CVGS was first introduced in~\cite{PhysRevA.73.032318} as the
continuous analogue of discrete variable graph state.  It has the nice
property that any local Gaussian operation on a CVGS can be associated
with a geometric transformation on its graph
representation~\cite{PhysRevA.78.052307}.
Refs.~\cite{PhysRevLett.97.110501, PhysRevA.79.062318} showed that
CVGS can be used to generate universal quantum operations and thus
is potentially a useful physical resource to implement quantum
computations. In addition, CVGS also finds applications in quantum
communications, {\it e.g.} Ref.~\cite{PhysRevA.78.042302} proposed a
protocol to realize quantum teleportation between two parties.

This paper is focused on the implementation and property analysis of
QSS schemes with CVGS. We differentiate eight QSS schemes according to
the secret and communication channel types. Among all these schemes,
four of them have no practical values because they are either
physically infeasible or insecure. We will thus investigate in the
other four schemes.  These extend the works
in~\cite{PhysRevA.78.042309, PhysRevA.82.062315} into the CVGS domain.

We study two essential problems for these QSS schemes with CVGS First,
for each QSS scheme, we design an implementation protocol for the
dealer and players so that the players may collaborate to estimate the
secret. The mean and variance of the estimation error are derived
explicitly. Based on the error statistics, we can derive the parameter
settings for unbiased estimation.  Furthermore, the protocol
parameters can be tuned to minimize the error variance.  In the case
of infinite squeezing, it can be shown that finding the condition that
a set of players can perfectly estimate the secret can be transformed
to solve a set of linear equations.

The second problem is the threshold protocol, which is crucial for
many applications that need decision-making. In QSS, a $(k,n)$
threshold protocol refers to the case when $k$ players or more can
estimate the secret perfectly, and any set with less players can never
get the secret within a finite error bound. We show that an arbitrary
$(k,n)$ threshold protocol with $n/2< k\le n$ can be implemented for
three schemes using a weighted CVGS prepared with infinitely squeezed
qumodes.  An interesting observation is that for the scheme with
quantum secret, private distribution channel, and quantum
player-player channel (referred as QPrtQ), the threshold protocol for
two non-cooperative player groups is exclusive, meaning that if one
group can perfectly estimate the secret qumode, the other cannot
estimate either quadrature of the secret qumode within a finite error
bound.  The security of the quantum secret is thus gurantteed.  For
QPrtQ and another scheme, these protocols cover all the physically
feasible cases, and we also reveal the duality between them.

This paper is organized as follows. Sec.~\ref{sec:background} provides
a brief introduction on QSS schemes and CVGS. Three QSS schemes with
CVGS are investigated in Sec.~\ref{sec:study-cplow-scheme},
\ref{sec:study-qplow-scheme} and~\ref{sec:study-cplow-schem},
respectively. We conclude the paper in Sec.~\ref{sec:conclusion}.

\section{Background}
\label{sec:background}
In this section we give a brief introduction of QSS schemes and CVGS.

\begin{figure}[bt]
 \includegraphics[width=0.95\hsize]{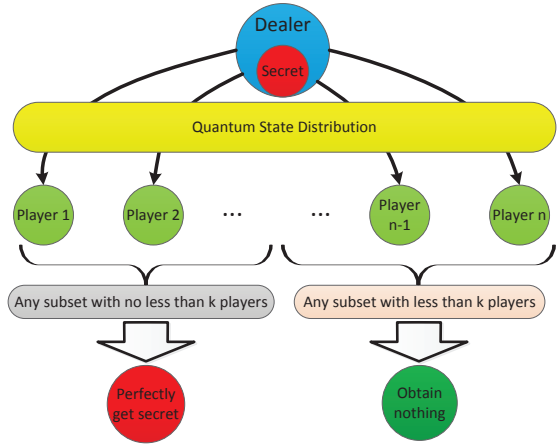}
  \caption{QSS  $(k,n)$ threshold protocol.}
 \label{fig3}
\end{figure}

In a QSS protocol, there are one dealer and $n$ players as shown in
Fig.~\ref{fig3}. The dealer has a secret that is represented by either
classical or quantum information.  At first, the dealer encodes the
secret into a prepared quantum state, and subsequently distributes it
to all the players through either private or public channels. With
this quantum state at hand, a group of players can either apply local
operations to their own states and then exchange classical
information, or take joint operations to their states.  The task for
these players is to reconstruct the secret based on the information
circulated around.

We can classify QSS into eight schemes according to their secret type
(classical or quantum), dealer-player distribution channel (private or
public), and player-player communication channel (classical or
quantum). Among all these eight schemes, QPubC and QPrtC are
physically infeasible because it is impossible to recover unknown
quantum information from classical information. Moreover, QPubQ and
CPubQ are insecure because an eavesdropper can disguise identity to
modify the information on public channel. Therefore, we will
investigate only the four schemes in Table~\ref{table:1}.

\begin{center}
\begin{table}[htbp]
\begin{tabular}[t]{p{40pt}|c|c|c}
\hline
\centering{} & Secret type & \tabincell{c}
{Dealer-Player \\Channel}
&\tabincell{c}{Player-Player \\Channel}  \\ \hline
\centering{CPvtC} &Classical &Private & Classical \\ \hline
\centering{QPvtQ} &Quantum&Private& Quantum\\ \hline
\centering{CPubC} &Classical &Public&Classical \\\hline
\centering{CPvtQ} &Classical &Private&Quantum \\\hline
\end{tabular}
\caption{Feasible QSS schemes.}
\label{table:1}
\end{table}
\end{center}

In particular, we are interested in a $(k,n)$ threshold protocol,
which refers to the case when it requires at least $k$ players to
estimate the secret perfectly, and any set with less than $k$ players
cannot estimate the secret within a finite error bound. This procedure
is illustrated in Fig.~\ref{fig3}.

In this paper we will use CVGS to implement QSS schemes.  A CVGS is an
entangled multi-qumode state that can be represented by an undirected
graph.  Denote the adjacency matrix of this graph as $G$, whose
element $G_{ij}$ represents the interaction gain of the coupling
between qumode $i$ and $j$. If $G_{ij}$ takes only binary values $0$
or $1$, it is an unweighted CVGS; otherwise, it is a weighted CVGS.

In a QSS scheme, the dealer needs to prepare a CVGS and then to encode
the secret into that CVGS. At the beginning, the dealer has $n$ vacuum
states each with the position $X_i^{(0)}$ and momentum $P_i^{(0)}$,
where both $X_i^{(0)}$ and $P_i^{(0)}$ are random variables with
standard Gaussian distribution. The dealer then squeezes the momentum
and at the same time amplifies the position of each qumode, obtaining
squeezed vacuum states:
\begin{equation}
  \label{eq:2}
  P_j=e^{-r_j}P_j^{(0)}, \quad
  X_j=e^{r_j}X_j^{(0)}.
\end{equation}
Here $r_j$ is the squeezing parameter for qumode $j$. Juxtapose
$X_j$'s and $P_j$'s in a vector form:
\begin{equation}
  \label{eq:4}
v_{(n)}=\begin{bmatrix}X_1&\cdots&X_n&P_1&\cdots&P_n \end{bmatrix}^T,
\end{equation}
where the subscript $(n)$ indicates the number of the qumodes.  Now
apply a quantum nondemolition (QND) coupling with interaction gain
$G_{ij}$ to the pair $(i, j)$~\cite{PhysRevA.73.032318}. This
establishes a connection between qumode $i$ and $j$ in the graph, and
the resulting quadratures are $(X_i,P_i+ G_{ij}X_j)$ and $(X_j, P_j+
G_{ji}X_i)$, respectively.  After a series of such QND coupling
operations, the final quadratures can be written as
\begin{equation}
\label{eq1}
X_j^G=X_j, \quad P_j^G=P_j+\sum_{l=1}^n G_{jl}X_l.
\end{equation}
Letting
\begin{equation*}
v_{(n)}^G=\begin{bmatrix}
X_1^G&\cdots&X_n^G& P_1^G&\cdots&P_n^G,
\end{bmatrix}^T.
\end{equation*}
we can rewrite \Eq{eq1} in a compact form as
\begin{equation}
  \label{eq:1}
  v_{(n)}^G=\ma{ I & 0 \\
    G_{(n)} & I} v_{(n)}.
\end{equation}

In the next three sections, we will investigate the implementations of
the CPvtC, QPvtQ, and CPubC schemes in Table~\ref{table:1}. We point
out that the CPvtQ scheme can be implemented by super-dense
coding~\cite{PhysRevLett.69.2881} and is indeed a quantum data hiding
scheme~\cite{985948, PhysRevLett.89.097905}. Since CPvtQ can be dealt
with similarly to the others, we will focus on the first three.  For
simplicity, we set $\hbar=1$ throughout this paper.

\section{Case 1: CP\lowercase{vt}C Scheme}
\label{sec:study-cplow-scheme}
In this section we study the CPvtC scheme, in which the dealer encodes
a {\it classical} secret into a CVGS, then distributes the qumodes to
the players through {\it private} channels, and finally the players
exchange information via {\it classical} channels so as to reconstruct
the secret. We will derive the estimation error and then obtain its
mean and variance. This facilitates the unbiased estimation and also
the optimal tuning of protocol parameters to minimize the error
variance. We will also study the condition to perfectly reconstruct
the secret, and discuss the implementation of a general threshold
scheme on CVGS.

We now present the implementation details of CPvtC scheme.  Assume
that the classical secret the dealer holds is a real number $\gamma$.
The dealer starts from encoding the secret into a CVGS by applying a
momentum displacement operation $Z(c_j \gamma) =e^{i c_j \gamma
  \hat{x}}$~\cite{PhysRevA.79.062318} to qumode $j$ with quadratures
$(X_j^G, P_j^G)$, where $c_j$, $\gamma$ are real numbers and $\hat{x}$
is the position operator. The momentum of qumode $j$ is shifted to
$P_j^G+c_j\gamma$. Let $c =\left[\,c_1\ \cdots\ c_n \right]^T$. Then
the shifted momenta for all the qumodes can be written as a vector
$c\gamma$. The dealer distributes qumode $j$ to player $j$ and
publishes the vector $c$ to all the players.  Now player $j$ has the
quadratures $(X_j^G, P_j^G+c_j\gamma)$ under disposal.

To recover the secret, player $j$ can take the following actions:
\begin{enumerate}
\item Let
  \begin{equation}
    \label{eq:3}
    P_j^D=P_j^G+c_j\gamma.
  \end{equation}
  Apply the operator $\exp\left\{-i\frac{\beta_j}{2\alpha_j}
    (\hat{P}_j^D)^2\right\}$ to the quadratures $(X_j,P_j^D)$ so that
  the new quadratures are $\left(X_j+\frac{\beta_j}{\alpha_j}P_j^D,
    P_j^D\right)$.

\item Measure the position to get $\CalM\left(X_j +
    \frac{\beta_j}{\alpha_j } P_j^D\right)$, where $\CalM(\,\cdot\,)$
  is a measurement operation that results in a random variable.

\item Scale the measurement result by $\alpha_j$ and obtain
\begin{equation}
\label{eq:11}
\begin{aligned}
  \mu_j&=\alpha_j\CalM\left(X_j+\frac{\beta_j}{\alpha_j }
  P_j^D\right)\\
&=\CalM(\alpha_j X_j+\beta_j P_j^D),
\end{aligned}
\end{equation}
where the last equality is because $\CalM(\,\cdot\,)$ is a linear
operation.
\end{enumerate}

The players can then exchange their $\mu_j$ by classical
communications. We now show that each player can use the sum of
$\mu_j$ as an estimation of the secret $\gamma$.  From
\Eqs{eq:4}-\eqref{eq:11}, the estimation error $e$ can be calculated
as
\begin{align}
    \label{eq2} \notag
    e=&\sum_{j=1}^n\mu_j-\gamma\\ \notag
    =&\CalM\left(\left[ a^T\mid b^T\right]
      \left(\left[\begin{array}{c|c}
            I & 0 \\\hline
            G_{(n)} & I
    \end{array}\right]
    v_{(n)}
    +\left[\begin{array}{cc}
    \bm{0} \\\hline
    c
    \end{array}\right]\gamma\right)\right) - \gamma\\
    =&\CalM\left(\left[a^T+b^TG_{(n)} \mid b^T\right]v_{(n)}\right)+(b^Tc- 1)\gamma,
\end{align}
where $a=[\,\alpha_1\ \cdots\ \alpha_n]^T$, $b=[\,\beta_1\ \cdots\
\beta_n]^T$, $\bm{0}=[\,0\ \cdots\ 0\,]^T$, and $G_{(n)}$ is the
adjacency matrix of the $n$-qumode graph state.

The mean of the estimation error is
\begin{equation*}
  \E e=\E \CalM\left(\left[a^T+b^TG_{(n)} \mid b^T\right]v_{(n)}\right)+(b^Tc- 1)\gamma.
\end{equation*}
Since $\E {\cal M}(X_j)=\E {\cal M}(P_j)=0$, we have $ \E
\CalM\left(\left[a^T+b^TG_{(n)} \mid b^T\right]v_{(n)}\right)=0$.
Hence,
\begin{displaymath}
 \E e=(b^Tc- 1)\gamma.
\end{displaymath}
To ensure an unbiased estimation, it is only required that
\begin{equation}
\label{eq7}
b^Tc=1.
\end{equation}
The variance of the estimation error can be obtained after some
algebraic derivations as
\begin{equation}
\label{eq4}
\Var(e)=\|(a^T+b^TG_{(n)})R_{(n)}\|^2+\|b^TR_{(n)}^{-1}\|^2,
\end{equation}
where $R_{(n)}=\diag\{e^{r_1},\cdots,e^{r_n}\}$, and $\|\cdot\|$ is the
Euclidean norm.

To enhance the estimation precision, it is desired to reduce the error
variance~\eqref{eq4}. Combining with \Eq{eq7}, it is a nonlinear
constrained minimization problem with optimization variables $a$, $b$,
$G_{(n)}$, and $R_{(n)}$. We can thus tune these protocol parameters
to achieve a better estimation. For example, when $a$, $b$, $G_{(n)}$
are fixed, the optimal squeezing parameters that minimize the error
variance can be chosen as
\begin{equation}
  \label{eq:13}
  r_j=\frac12 \log \left| \frac{b_j}{(a^T+b^T G_{(n)})_j}\right|,
\end{equation}
 for $b_j\neq 0$ and $(a^T+b^T G_{(n)})_j \neq 0$.
 
 It is also easy to observe that under constraint~\eqref{eq7}, the
 variance~\eqref{eq4} can achieve 0 only if some parameters take
 extremal values. One choice is to apply infinite squeezing, {\it
   i.e.}  letting the squeezing parameters $r_j\to \infty$ and then
\begin{equation}
\label{eq6}
a^T+b^T G_{(n)}=\bm{0}^T.
\end{equation}
Combining \Eqs{eq7} and \eqref{eq6}, we get a condition that
guarantees $n$ players to get the secret perfectly under infinite
squeezing
\begin{equation}
    \label{eq8}
    \left[a^T\mid b^T  \right]
    \left[\begin{array}{c|c}
    I & \bm{0}\\\hline
    G_{(n)} & c
    \end{array}\right]
    =\left[\bm{0}^T\mid 1\right].
\end{equation}

Now let us discuss $(k, n)$ QSS threshold protocols, which means that
it requires at least $k$ ($k\le n$) players to estimate the secret
perfectly, and any set with less than $k$ players cannot estimate the
secret within a finite error bound. Consider a set of $k$
collaborative players with indices $j_1$, $\cdots$, $j_k$.  To
simplify the notation, we use $A_{J,K}$ to denote a matrix formed by
taking rows with indices in $J$ and columns in $K$ from a matrix $A$,
where $J, K$ are subsets of $N=\{1,\cdots,n\}$.  For the case of a
vector, we can similarly define $v_J$. Removing the rows and columns
corresponding to the remaining $n-k$ players from \Eq{eq8}, we obtain
\begin{equation}
\label{eq10}
\left[\begin{array}{cc}
    a^T_J & b^T_J
\end{array}\right]
\left[\begin{array}{cc}
    I_{J,N} & \bm{0} \\
    G_{J,N} & c_{J}  \\
    \end{array}\right]
    =\begin{bmatrix}0&\cdots&0&1\end{bmatrix},
\end{equation}
where $J=\{j_1,\cdots,j_k\}$.  \Eq{eq10} is a sufficient and necessary
condition for $k$ players from a set of $n$ players to recover the
secret perfectly under infinite squeezing.

We give a lower bound on $k$ that ensures the physical existence of a
$(k, n)$ CPvtC threshold protocol.

\begin{theorem}
\label{thm:1}
A $(k,n)$ threshold protocol of CPvtC scheme satisfying $n/2 < k\le n$
can be implemented on a weighted CVGS with infinite squeezing.
\end{theorem}
To keep the flow of the paper, the proof is given in
Appendix~\ref{Appendix:A}.

\section{Case 2: QP\lowercase{vt}Q scheme}
\label{sec:study-qplow-scheme}
In this section we discuss the QPvtQ scheme, in which the dealer has a
{\it quantum} secret, the qumodes encoding the secret are distributed
through {\it private} channels, and the players share their
information by {\it quantum} communication channels. We will first
give the implementation protocol design, and then calculate the
estimation error.  We then discuss the condition of perfectly
estimating the secret qumode as well as the threshold protocols under
infinite squeezing.

First consider the protocol design. In a QPvtQ scheme, the dealer has
a secret qumode $(X_S,P_S)$. At the beginning, the dealer prepares an
$(n+1)$-mode CVGS, and keeps the $(n+1)$-th qumode with quadratures
$(X_{n+1}^G, P_{n+1}^G)$ for later use. The dealer distributes the
other $n$ qumodes to the $n$ players. Now the dealer performs a Bell
measurement as follows.  First, combine the $(n+1)$-th qumode with
$(X_S,P_S)$ to yield two new qumodes $(X_u,P_u)$ and $(X_v,P_v)$,
where
\begin{align}\notag
 &X_u=\frac{X_{n+1}^G+X_S}{\sqrt{2}},\qquad P_u=\frac{P_{n+1}^G+P_S}{\sqrt{2}}\\
 \label{eq16}
 &X_v=\frac{X_{n+1}^G-X_S}{\sqrt{2}},\qquad P_v=\frac{P_{n+1}^G-P_S}{\sqrt{2}}.
\end{align}
Second, take homodyne measurements for $X_u$ and $P_v$. The
measurement results $\CalM(X_u)$ and $\CalM(P_v)$ are two Gaussian
random variables.

The dealer publishes these two measurement results to all the players.
If any set of players can construct the qumode
$(-X_{n+1}^G,P_{n+1}^G)$, they can perfectly estimate the secret by
simply adding the position displacement $\sqrt{2}\CalM(X_u)$ and
subtracting the momentum displacement
$\sqrt{2}\CalM(P_v)$~\cite{RevModPhys.77.513}. This is the idea of
continuous variable quantum teleportation~\cite{PhysRevLett.80.869}.

To construct $(-X_{n+1}^G,P_{n+1}^G)$, the players can take the
following steps:
\begin{enumerate}
\item Apply a single-mode Gaussian unitary operation and a phase
  insensitive amplification~\cite{PhysRevA.83.052307} to transform a
  qumode $(X_j^G,P_j^G)$ to $(\alpha_j X_j^G+ \beta_j P_j^G, \alpha'_j
  X_j^G+ \beta'_j P_j^G)$, where $\alpha_j$, $\beta_j$, $\alpha'_j$,
  $\beta'_j$ are all real numbers;
  
\item Pick one qumode from the players' qumodes and transform it to
  $(\sum_{i=1}^n\alpha_jX_j^G+\beta_j
  P_j^G,\sum_{i=1}^n\alpha'_jX_j^G+\beta'_j P_j^G)$ by using nonlocal
  operations such as a controlled-X  operation~\cite{PhysRevA.81.022311}.
\end{enumerate}

From \Eq{eq:1}, the position error can be calculated as
\begin{align}\notag
e_x=&\sum_{i=1}^n(\alpha_jX_j^G+\beta_j P_j^G)-(-X_{n+1}^G)\\ \notag
    =&\left[a^T\ 0\ b^T\ 0\right]
    \left[\begin{array}{c|c}
    I & 0 \\\hline
    G_{(n+1)} & I
    \end{array}\right] v_{(n+1)}\\ \notag
    &+
    [\bm{0}^T_{(n)}\ 1\ \bm{0}^T_{(n+1)}]v_{(n+1)} \\
=&\left[\begin{array}{c|c}[a^T\, 1]+[b^T\,
  0]G_{(n+1)}&\, [b^T \,0]\end{array}\right]v_{(n+1)},
  \label{eq:9}
\end{align}
where $a=[\alpha_1, \cdots, \alpha_n]^T$, $b=[\beta_1, \cdots,
\beta_n]^T$, $v_{(n+1)}=[X_1, \cdots, X_{n+1}, P_1, \cdots,
P_{n+1}]^T$, and $G_{(n+1)}$ is an $(n+1)\times(n+1)$ adjacency
matrix.  Similarly, the momentum error is
\begin{align}\notag
    e_p=&\sum_{i=1}^n(\alpha'_jX_j^G+\beta'_j P_j^G)-P_{n+1}^G\\ \notag
    =&\left[a'^T\ 0\ b'^T\ 0\right]
    \left[\begin{array}{c|c}
    I & 0 \\\hline
    G_{(n+1)} & I
    \end{array}\right] v_{(n+1)}\\ \notag
  &-[g_{n+1}^T\ \bm{0}^T_{(n)}\ 1]v_{(n+1)} \\
  =&\left[\begin{array}{c|c}[a'^T\, 0]+[b'^T\,0]G_{(n+1)}-g_{n+1}^T
      &\, [b'^T\, -\!1]\end{array}\right]v_{(n+1)},
    \label{eq5}
\end{align}
where $a'=[\alpha'_1, \cdots, \alpha'_n]^T$, $b'=[\beta'_1, \cdots,
\beta'_n]^T$, and $g_{n+1}^T$ is the $(n+1)$-th row of the matrix
$G_{(n+1)}.$

By applying local unitary operations, the covariance matrix of the
secret qumode can be diagonalized to
\begin{equation*}
\left(  \begin{matrix}
     \Var(X_S) & 0 \\
    0 & \Var(P_S)
  \end{matrix}\right)
\end{equation*}
From  Eq.~(1) in~\cite{jing2011dependence}, we
can get the fidelity of the estimated secret qumode as
\begin{equation}
\label{eq15}
F=\frac{2}{\sqrt{\delta+\epsilon}-\sqrt{\epsilon}},
\end{equation}
where
\begin{align*}
\delta=&(2\Var(X_S)+V_1)(2\Var(P_S)+V_2), \\
\epsilon=&(\Var(X_S)\Var(P_S)-1)\times\\
 &\quad[(\Var(X_S)+V_1)(\Var(P_S)+V_2)-1], \\
V_1=&\left\|\left[[a^T\, 1]+[b^T\, 0]G_{(n+1)}\right]R_{(n+1)}\right\|^2 \\
   &+\left\|[b^T\, 0]R_{(n+1)}^{-1}\right\|^2, \\
V_2=&\left\|\left[[a'^T\, 0]+[b'^T\, 0]G_{(n+1)}-g_{n+1}^T\right]R_{(n+1)}\right\|^2\\
   &+\left\|[b'^T\,-\!\!1]R_{(n+1)}^{-1}\right\|^2, \\
R_{(n+1)}=&\diag\{e^{r_1},\cdots,e^{r_{n+1}}\}.
\end{align*}
In particular, for minimum uncertainty states, we have that $
\Var(X_S)\Var(P_S)=1$. Hence $\epsilon=0$, and~\Eq{eq15} can be
simplified to
\begin{equation}
  \label{eq:14}
    F=\frac{2}{\sqrt{\delta}}.
\end{equation}


With the fidelity in \Eq{eq15}, it is possible to optimize the
protocol parameters to maximize the fidelity. 
To achieve perfect fidelity at 100\%, it is required that $V_1=V_2=0$.
This amounts to the following conditions under infinite squeezing:
\begin{align}
\label{eq22}
&\left[[a^T\, 1]+[b^T\, 0]G_{(n+1)}\right]=\bm{0}^T, \\
\label{eq23}
&\left[[a'^T\, 0]+[b'^T\, 0]G_{(n+1)}-g_{n+1}^T\right]=\bm{0}^T.
\end{align}
\Eqs{eq22} and~\eqref{eq23} can be rewritten as 
 \begin{align}
   \label{eq17}
    & \left[a^T\mid b^T\right]
    \left[\begin{array}{c}
    I' \\\hline
    G_{(n+1)}'
    \end{array}\right]
    =\left[\bm{0}^T\mid -1\right], \\
    \label{eq18}
    &  \left[a'^T\mid b'^T\right]
    \left[\begin{array}{c}
    I' \\\hline
    G_{(n+1)}'
    \end{array}\right]
    = g_{n+1}^T,
\end{align}
where $I'$, $G_{(n+1)}'$ are $n\times (n+1)$ matrices obtained by
deleting the $(n+1)$-th row of the matrices $I$ and $G_{(n+1)}$,
respectively.

Next we study the threshold protocol for QPvtQ scheme. The following
theorem can be obtained.
\begin{theorem}
\label{thm:3}
  Any $(k,n)$ threshold protocol of QPvtQ scheme can be implemented
  with a weighted CVGS of infinite squeezing.
\end{theorem}
The proof is given in Appendix~\ref{Appendix:C}.

Furthermore, different from CPvtC, if these $k$ players can perfectly
recover the secret, we can show that the remaining $n-k$ players
cannot get any information about the secret.

\begin{theorem}
\label{thm:2}
For two non-cooperative group with QPvtQ scheme, if one group can
perfectly estimate the secret qumode, the other group cannot estimate
either quadrature of the quantum secret within a finite error bound.
Thus they cannot obtain any information about the secret.
\end{theorem}
The proof is provided in Appendix~\ref{Appendix:B}.

For a $(k,2k-1)$ threshold protocol, since any group with $k$ or more
players can perfectly estimate the secret, from Theorem~\ref{thm:2},
we know that any group with less than $k$ players can obtain no
information about the quantum secret.  This holds true for any $(k,n)$
threshold protocol, which is obtained from $(k,2k-1)$ protocol by
picking $n$ qumodes from $2k-1$ qumodes. For these protocols, we have
the following corollary.
\begin{corollary}
\label{cor1}
Any player group with number less than the threshold $k$ cannot obtain any
information about the quantum secret.
\end{corollary}

\section{Case 3: CP\lowercase{ub}C scheme}
\label{sec:study-cplow-schem}
This section is focused on the CPubC scheme, where the dealer has a
{\it classical} secret, the qumodes encoding this secret is
distributed through {\it public} channels, and the players collaborate
to get the secret by {\it classical} communication channels. We will
propose an implementation protocol, and then calculate the estimation
error. The threshold protocol is studied by revealing the duality
between QPvtQ and CPubC and schemes.

We start from proposing the implementation protocol. First, the
dealer prepares an $(n+1)$-mode CVGS, keeps the $(n+1)$-th qumode, and
then distributes the other $n$ qumodes to the $n$ players.  Since the
qumodes are distributed through public channels, there exists risk
that some eavesdroppers may get them. To ensure secure classical
communications, from the method of CV quantum key
distribution~\cite{bennett1984quantum}, the dealer takes a random
homodyne measurement at the $(n+1)$-th qumode and obtains either
$\CalM(X_{n+1}^G)$ or $\CalM(P_{n+1}^G)$. Here the dealer measures
either the position or the momentum, but which quadrature has been
measured is unknown to the others.  The measurement outcome is then
used as a random key that the dealer will share with the players.

Secondly, the players achieves a consensus via classical communications
that they will randomly estimate either $\CalM(X_{n+1}^G)$ or
$\CalM(P_{n+1}^G)$ in a collaborative manner.  Then, they take the
three steps of \Eqs{eq:3}-\eqref{eq:11} as in
Sec.~\ref{sec:study-cplow-scheme}, and exchange their results so as to
use $\sum_{j=1}^n\CalM(\alpha_jX_j+\beta_jP_j^G)$ as an estimation of
the secret.

Thirdly, both the dealer and the players need to make sure that
  the quadrature they estimated is exactly the same as the one that
  the dealer measured earlier. The dealer and the players will do the
  following:
\begin{enumerate}
\item The players announce the quadrature that they estimated;
\item The dealer publishes the quadrature actually measured;
\item If the quadrature estimated by the players matches the one
  measured by the dealer, they keep the estimation result
  $\sum_{j=1}^n\CalM(\alpha_jX_j+\beta_jP_j^G)$ as the shared key; if
  not, they discard it and try again.
\end{enumerate}
Step $3$ is necessary because if the estimation quadrature matches the
measurement quadrature, the players obtain an unbiased estimation of
the measurement outcome. Otherwise, the players get something
completely useless. The error in this case will be unbounded, as a
homodyne measurement for the position (or momentum) will collapse the
momentum (or position) into a maximally uncertain state. This
completes the protocol implementation.

Next we calculate the estimation errors for both quadratures.  If the
players have estimated $\CalM(X_{n+1}^G)$, the position estimation
error is
\begin{align}
\label{eq12}
  e_x=&\CalM\left(\left[a^T\ 0\ b^T\ 0\right]
    \left[\begin{array}{c|c}
    I & 0 \\\hline
    G_{(n+1)} & I
    \end{array}\right] v_{(n+1)} \right) \nonumber -\CalM\left(X_{n+1}^G\right) \notag \\
 =&\CalM\left(\left[\begin{array}{c|c}[a^T\, -\!1]+[b^T\,
         0]G_{(n+1)}&[b^T\,0]\end{array}\right]v_{(n+1)}\right).
\end{align}
It is easy to see that the error has zero mean and we have an unbiased
estimation. The error variance is given by
\begin{align} \notag
\Var(e_x)=&
\left\|\left[[a^T\, -\!1]+[b^T\,
         0]G_{(n+1)}\right]R'\right\|^2 \\
         &+\left\|[b^T\,
         0]R'^{-1}\right\|^2.
         \label{eq19}
\end{align}
The variance achieves $0$ only when the qumodes are infinitely
squeezed and the following equation holds true:
\begin{equation}
\label{eq13}
[a^T\, -\!1]+[b^T\,0]G_{(n+1)}=\bm{0}^T.
\end{equation}
\Eq{eq13} can be rewritten as
\begin{equation}
\label{eq14}
    [a^T\mid b^T]
    \left[\begin{array}{c}
    I' \\\hline
    G_{(n+1)}'
    \end{array}\right]
    =[\bm{0}^T\mid 1].
\end{equation}

If the players have estimated $\CalM(P_{n+1}^G)$, the momentum
estimation error is
\begin{align}
  &e_p\\
  =&\CalM\left(\left[a'^T\ 0\ b'^T\ 0\right]
    \left[\begin{array}{c|c}
    I & 0 \\\hline
    G_{(n+1)} & I
    \end{array}\right] v_{(n+1)} \right) \nonumber -\CalM\left(P_{n+1}^G\right) \notag \\
 =&\CalM\left(\left[\begin{array}{c|c}[a'^T\, 0]+[b'^T\,
         0]G_{(n+1)}-g_{n+1}^T&[b'^T\,-\!1]\end{array}\right]v_{(n+1)}\right).
\end{align}
The error $e_p$ also has zero mean and we again have an unbiased
estimation. Its variance is given by
\begin{align} \notag
\Var(e_p)=&\left\|\left[[a'^T\, 0]+[b'^T\,
         0]G'-g_{n+1}^T\right]R'\right\|^2\\
         &+\left\|[b'^T\,-\!1]R'^{-1}\right\|^2.
\label{eq:12}
\end{align}
To make the error variance equal to $0$, we need the infinite
squeezing together with
\begin{equation}
\left[[a'^T\, 0]+[b'^T\,
         0]G'-g_{n+1}^T\right]=\bm{0}^T,
         \label{eq24}
\end{equation}
which yields that
\begin{equation}
\label{eq3}
[a'^T\mid b'^T]
 \left[\begin{array}{c}
    I' \\\hline     G''
    \end{array}\right]
    =g_{n+1}^T.
\end{equation}

Finally, we discuss the threshold protocol of CPubC by revealing the
duality between QPvtQ and CPubC schemes. We now show that under
infinite squeezing, a $(k,n)$ threshold protocol can be implemented on
CPubC if and only if it can be implemented on QPvtQ.  We have proved
that in CPubC scheme the existence of a set of players who can
perfectly estimate the secret is equivalent to the consistency of
\Eqs{eq14} and~\eqref{eq3}, and in QPvtQ scheme that existence is
equivalent to the consistency of \Eqs{eq17} and~\eqref{eq18}. It is
clear that \Eqs{eq18} and~\eqref{eq3} are the same, and \Eq{eq17}
differs from \Eq{eq14} only by a sign. Thus the existence of a $(k,n)$
threshold protocol on CPubC is equivalent to that on QPvtQ.  Similar
results for the discrete variable were given in~\cite{1205.4182}.
Furthermore, from Theorem~\ref{thm:3}, a $(k,n)$ threshold CPubC
protocol exists if and only if $n/2 < k\le n$, and all these CPubC
protocols can be implemented using weighted CVGSs.

\section{conclusion}
\label{sec:conclusion}
This paper investigated three QSS schemes with CVGS in details,
namely, CPvtC, QPvtQ, and CPubC. We designed implementation protocols
for each scheme, and derived analytic formula for the estimation
error. This makes it possible to minimize the error variance by
varying protocol parameters. We also showed that a $(k,n)$ threshold
QSS protocol of the three schemes satisfying $n/2< k\le n$ can be
implemented by using a weighted CVGS with infinite squeezing. These
protocols cover all the physically feasible threshold protocols for
QPvtQ and CPubC. Specifically, the perfect estimation for two
non-cooperative groups on QPvtQ is exclusive.  Finally, the duality
between QPvtQ and CPubC schemes is discussed.


\begin{acknowledgments}
  JZ thanks the financial support from the Innovation Program of
  Shanghai Municipal Education Commission under Grant No. 11ZZ20,
  Shanghai Pujiang Program under Grant No. 11PJ1405800, NSFC under
  Grant No. 61174086, Project-sponsored by SRF for ROCS SEM, and State
  Key Lab of Advanced Optical Communication Systems and Networks,
  SJTU, China. GQH thanks the financial support from NSFC under Grant
  No. 61102053, Project-sponsored by SRF for ROCS SEM, and SMC
  Excellent Young Faculty Award.
\end{acknowledgments}

\appendix

\section{Proof of Theorem~\ref{thm:1}}
\label{Appendix:A}
To guarantee all the $(k,n)$ threshold protocols with $n/2< k\le n$ can
be implemented, the dealer only need to make sure that they can
implement the case when $n=2k-1$. In $(k,2k-1)$ threshold protocols,
any $k$ players can cooperatively get the secret. Even if less than
$k$ of the $2k-1$ qumodes are removed, any $k$ players holding the
reserved qumodes can still obtain the secret. Hence, by choosing
arbitrary $n$ players from the total $2k-1$ players, a $(k,2k-1)$
threshold protocol can be transformed into a $(k,n)$ protocol.  Thus,
to prove Theorem~\ref{thm:1}, we only need to show that any $(k,2k-1)$
protocol can be implemented using a weighted CVGS of infinite
squeezing.

Suppose that in a communication system with one dealer and $2k-1$
players, a set of $k$ players collaborate to reveal the secret. Since
\Eq{eq10} is a sufficient and necessary condition for the $k$ players
to perfectly estimate the secret, to guarantee they can get the
secret, it is required that \Eq{eq10} with $n=2k-1$ has solutions. In
\Eq{eq10}, the $2k\times 2k$ matrix
\[
\begin{bmatrix}
  I_{J,N}&\bm{0} \\
  G_{J,N}&c_{J}
\end{bmatrix}
\]
maps a $2k$-dimensional vector $[a_{J}^T\ b_{J}^T]^T$ to a
$2k$-dimensional nonzero vector $[0\,\cdots\,0\,1]^T$, where
$J=\{j_1,\cdots,j_k\}$ and $N=\{1,\cdots,2k-1 \}$. If this matrix is
full rank, there exists exactly one solution $[a_J^T\ b_J^T]$.  Since
the submatrix $I_{J,N}$ is always full rank, we only need to guarantee
the submatrix $[G_{J,K}\ c_J]$ is full rank, where $K=N\setminus J$.
This condition can be satisfied by designing the adjacency matrix $G$
and the vector $c$. Here the backslash denotes the set difference.

To show that it is a $(k,2k-1)$ threshold protocol, we also need to prove
that any subset with fewer than $k$ players cannot estimate the secret
within a finite error bound. Indeed, we only need to prove there is no
solution to \Eq{eq10} if $k$ is replaced by $k-1$. In this case,
\Eq{eq10} becomes
\begin{align}
\begin{bmatrix}
  a_{J'}^T& b_{J'}^T
\end{bmatrix}
\begin{bmatrix}
  I_{J',N}&\bm{0} \\
  G_{J',N}&c_{J'}
\end{bmatrix}=
\begin{bmatrix}
  0&\cdots&0&1
\end{bmatrix},
  \label{eq:5}
\end{align}
where $J=\{j'_1,\cdots,j'_{k-1}\}$.
Consider the first $2k-1$ columns of the matrix in  \Eq{eq:5}. The submatrix
\[
\begin{bmatrix}
  I_{J',N}\\
  G_{J',N}
\end{bmatrix}
\]
maps $[a_{J'}^T\ b_{J'}^T]$ to a ($2k-1$)-dimensional zero vector.
Since the submatrix is full rank, $[a_{J'}^T\ b_{J'}^T]$ can only be a
zero vector, which contradicts the fact that $b_{J'}^Tc_{J'}=1$. So
\Eq{eq:5} has no solutions. Hence the theorem is proved.

\section{Proof of Theorem~\ref{thm:3}}
\label{Appendix:C}
From quantum no-cloning theorem, we know that a $(k,n)$ threshold
QPvtQ protocol must satisfy $n/2<k\le n$.  The largest possible value
of $n$ is $2k-1$. In this case, $\left[ I'^T \mid G'^T\right]^T$ is a
$2n\times (n+1)$ matrix. Since there are $2(n-k)$ zeros in $[\,a^T\mid
b^T]$, only a $2k\times 2k$ submatrix $[(I_J)^T \ (G_{J,N})^T]^T$ needs to
be considered in \Eqs{eq17} and~\eqref{eq18}. If this matrix is full
rank, both \Eqs{eq17} and~\eqref{eq18} have a unique solution.  The
matrix $I_J$ is always full rank, thus to make $[(I_J)^T \ (G_{J,N})^T]^T$
full rank, we need to the $k\times k$ submatrix $G_{J,K}$ to be full
rank as well, where $K=N\setminus J$.

If for any $k$ players, the corresponding $G_{J,K}$ is full rank, this
CVGS can be used to implement a $(k,2k-1)$ threshold QPvtQ protocol.
We can always find a proper weighted CVGS satisfying this condition.
If $(k,2k-1)$ protocols are obtained, the dealer can implement any
$(k,n)$ protocol by picking $n$ qumodes from a $(2k-1)$-mode CVGS and
distributing to $n$ players.

\section{Proof of Theorem~\ref{thm:2}}
\label{Appendix:B}
Divide $n$ players into two groups: one has $k$ players and the other
$n-k$ players. We need to show that if one group can perfectly
estimate the secret qumode $(X_S,P_S)$, the other group cannot
estimate either $X_S$ or $P_S$ within a finite error bound. If we can
prove it is impossible that one group perfectly estimates $X_S$ when
the other group perfectly estimates $P_S$, the theorem is proved
because any nonzero estimation error must be unbounded under infinite
squeezing.

If the group with $k$ players can collaborate to estimate the position
distribution of the secret qumode perfectly, we have
 \begin{equation}
    \begin{bmatrix}
   a_J^T&b_J^T
 \end{bmatrix}
 \begin{bmatrix}
   I_{J,M}\\G_{J,M}
 \end{bmatrix}=
 \begin{bmatrix}
   \bm{0}^T_n&-1
 \end{bmatrix},
 \label{eq:10}
 \end{equation}
 where $M=\{1,\cdots,n+1\}$,  and $J$ is a $k-$subset of
 $N=\{1,\cdots,n\}$.  From \Eq{eq:10}, we obtain
\begin{equation}
  \label{eq:6}
  b_J^T G_{J,M\setminus J}=\begin{bmatrix}\bm{0}^T_{n-k}&-1\end{bmatrix}.
\end{equation}
Denote the last column of $G_{J,M\setminus J}$ as $v_1$.

For the other group, if they can collaborate to estimate the momentum
distribution of the secret mode, we get
\begin{equation}
  \label{eq:7}
  \begin{bmatrix}
    a^T_K&b_K^T
  \end{bmatrix}
  \begin{bmatrix}
    I_{K,M}\\G_{K,M}
  \end{bmatrix}=g_{n+1}^T,
\end{equation}
where $K=N\setminus J$. We then have
\begin{equation}
  \label{eq:8}
  b_K^TG_{K,P}=v_2^T,
\end{equation}
where $P=M\setminus K$ and $v_2=(g_{n+1})_P$ (recall that $g_{n+1}$ is
the last column of $G_{(n+1)}$).  Hence $v_2^T=[v_1^T\ 0]$. Since
$G_{J,N}=\left[G_{J,K}\ v_1\right]$, we can rewrite \Eqs{eq:6}
and~\eqref{eq:8} as
\begin{align}
\label{eq20}
    &b_J^T
    \begin{bmatrix}
      G_{J,K} & v_1
    \end{bmatrix}
=\begin{bmatrix}\bm{0}^T_{n-k}&-1\end{bmatrix}, \\
\label{eq21}
    &b_K^T
    \begin{bmatrix}
      G_{K,J} & v_3
    \end{bmatrix}
=\begin{bmatrix}v_1^T&0\end{bmatrix},
\end{align}
where $v_3$ is the last column of $G_{K,P}.$ From \Eq{eq21}, we have
$v_1^T=b_K^TG_{K,J}$.  Substituting it into \Eq{eq20}, we get
\begin{equation*}
b_J^TG_{J,K}[I\mid b_K]=[\bm{0}^T_{n-k}~-1],
\end{equation*}
which is a contradiction.  Thus, it is impossible for one group of players
to perfectly estimate the position distribution, and the other to
estimate the momentum distribution, if these two groups do not
have any quantum communication.

\bibliographystyle{apsrev}

\end{document}